\documentclass[pra,reprint,twocolumn,nofootinbib,superscriptaddress,nopacs,aps,10pt]{revtex4-1}
\usepackage[utf8]{inputenc}
\usepackage[english]{babel}
\usepackage[T1]{fontenc}
\usepackage{amsmath}
\usepackage{hyperref}
\usepackage{graphicx}
\usepackage{amsfonts}
\usepackage{mathrsfs} 
\usepackage{amsthm}
\usepackage{amsmath}
\usepackage{amssymb}
\usepackage{color}
\usepackage{bbm} 
\usepackage{bm}  
\usepackage{mathtools} 

\newcommand{\ket}[1]{\vert #1 \rangle}
\newcommand{\bra}[1]{\langle #1 \vert}

\newcommand\proj[1]{\vert #1 \rangle \langle #1 \vert}

\newcommand{\opn}[1]{\operatorname{#1}}
\DeclareMathOperator{\tr}{Tr}  
\newcommand{\renyi}{R\'enyi~}
\newcommand*{\bR}{\mathbb{R}}

\newcommand*{\bX}{\mathbb{X}}

\newcommand*{\bZ}{\mathbb{Z}}
\newcommand*{\cH}{\mathcal{H}}
\newcommand*{\cI}{\mathcal{I}}
\newcommand*{\cJ}{\mathcal{J}}
\newcommand*{\cS}{\mathcal{S}}
\newcommand{\density}[1]{\mathscr{D}(#1)}
\newcommand{\operatorNorm}[1]{\left\lVert#1\right\rVert_{\infty}}
\newcommand{\SEP}{\opn{SEP}} 

\theoremstyle{example}

\newtheorem{theorem}{Theorem}
\newtheorem{lemma}[theorem]{Lemma}

\theoremstyle{definition}

\begin{document}

\title{Entanglement Detection via Direct-Sum Majorization Uncertainty Relations}
\author{Kun Wang}
\email{wk@smail.nju.edu.cn}
\author{Nan Wu}
\email[Corresponding author:~]{nwu@nju.edu.cn}
\author{Fangmin Song}
\email[Corresponding author:~]{fmsong@nju.edu.cn}
\affiliation{State Key Laboratory for Novel Software Technology,
            Department of Computer Science and Technology, Nanjing University, Nanjing 210093, China}
\date{\today}
\begin{abstract}
In this paper we investigate the relationship between direct-sum majorization formulation of uncertainty relations and
entanglement, for the case of two and many observables. Our primary results are entanglement detection methods based on
direct-sum majorization uncertainty relations. These nonlinear detectors provide a set of necessary conditions for
detecting entanglement whose number grows with the dimension of the state being detected. 
\end{abstract}

\maketitle

\section{Introduction}


Uncertainty relations form a central part of quantum mechanics. They impose fundamental limitations on our ability to
simultaneously predict the outcomes of noncommuting observables. Different approaches have been proposed to quantify
these relations. The original formulation is given by Heisenberg~\cite{heisenberg1927w} in terms of standard deviations
for momentum and position operators. His result is then generalized to two arbitrary
observables~\cite{robertson1929uncertainty}.
Later it is recognized that one can express uncertainty relations in terms of 
entropies~\cite{bialynicki1975uncertainty,deutsch1983uncertainty,maassen1988generalized}. 
In this approach, entropy functions like Shannon and \renyi entropies are
used to quantify uncertainty (Ref.~\cite{coles2017entropic} is a nice survey on this topic).
However, entropies are by no reason the most adequate to use. 
With this motivation, majorization is used to study uncertainty
relations~\cite{partovi2011majorization}. This line of research is further investigated
in~\cite{friedland2013universal,puchala2013majorization,rudnicki2014strong}.


Entanglement is another appealing feature of quantum mechanics and has been extensively investigated in the past
decades~\cite{horodecki2009quantum}. Entangled states play important roles in quantum information processing, such as
quantum teleportation~\cite{bennett1993teleporting} and dense coding~\cite{bennett1992communication}. Deciding whether
a given quantum state is entangled is a key problem of quantum information theory and known to be computationally
intractable in general~\cite{gurvits2004classical}. Therefore, computationally tractable necessary conditions for
entanglement detection, which provide a partial solution, have been the subject of active research in
recent years~\cite{guhne2009entanglement}.

Refs.~\cite{giovannetti2003characterizing,hofmann2003violation} present several methods for detecting
entanglement via variance based uncertainty relations. Similar methods have been designed using entropy based
uncertainty relations~\cite{giovannetti2004separability,guhne2004characterizing}. One may wonder whether there
exists a relationship between the majorization based uncertainty relation and entanglement. The answer is affirmative.
In~\cite{partovi2012entanglement}, the author applies the tensor-product majorization formulation of
uncertainty~\cite{partovi2011majorization} to the problem of entanglement detection. In this
paper we use the direct-sum majorization uncertainty relation, developed in~\cite{rudnicki2014strong}, to design an
entanglement detection method. As the direct-sum majorization bound has analytical solution
while the tensor-product majorization bound does not, 
our direct-sum majorization based detection method is more practical than the tensor-product majorization based method.

The rest of this paper is organized as follows. In Sec.~\ref{sec:direct-sum}, we
establish the notation and briefly review the direct-sum majorization formulation of uncertainty. 
In Sec.~\ref{sec:detection}, we present our central result --- an entanglement detection method 
based on the direct-sum majorization uncertainty.
In Sec.~\ref{sec:many-observables}, we generalize our result to the case of many observables.
We conclude in Sec.~\ref{sec:conclusion}. Some proofs are given in the Appendix.
\section{Direct-sum majorization uncertainty relations}\label{sec:direct-sum}

This section presents a basic review of the majorization theory and the 
formulation of direct-sum majorization approach to uncertainty relations.

\subsection{Majorization}\label{eq:majorization}

Let $\bR_+ = [0,\infty)$ be the set of non-negative real numbers,
$\bR_+^d = \{(p_1,\cdots, p_d):p_i \in \bR_+\}$ be the set
of $d$-dimensional real vectors with non-negative components. 
We denote by $\bm{p}\in\bR_+^d$ a $d$-dimensional vector and by $p_i$ the $i$-th element of $\bm{p}$.
For any vector $\bm{p}\in\bR_+^d$, let
$\bm{p}^\downarrow$ be the vector obtained from $\bm{p}$ by arranging the components of the latter
in descending order. Given two vectors $\bm{p}, \bm{q}\in\bR_+^d$, $\bm{p}$ is said to be
\textit{majorized} by $\bm{q}$ and written $\bm{p} \prec \bm{q}$ if
\[
\forall k \in [d-1],\; \sum_{i=1}^{k} p^\downarrow_i \leq \sum_{i=1}^{k} q^\downarrow_i,
\quad\text{and~} \sum_{i=1}^{d} p^\downarrow_i = \sum_{i=1}^{d} q^\downarrow_i,
\]
where $[d] =\{1,\cdots,d\}$.
Intuitively, $\bm{p} \prec \bm{q}$ means that the sum of largest $k$ components of $\bm{p}$
is no larger than the sum of $k$ largest components of $\bm{q}$. The majorization order is a partial order, i.e.,
not every two vectors are comparable under majorization.
When studying majorization among two vectors of different dimensions,
we append $0$(s) to the vector with smaller dimension so that two vectors have the same dimension.

A related concept is the \textit{supremum} of a set of $N$ vectors,
defined as the vector that majorizes every element of the set and, is majorized by any vector
that has the same property. We now briefly describe how to construct the supremum vector, 
more details can be found in~\cite{cicalese2002supermodularity,partovi2011majorization}.
Let $\cS=\{\bm{p}^{(1)}, \cdots, \bm{p}^{(N)}: \bm{p}^{(n)} \in \bR_+^d\}$ a set of $N$ vectors.
To construct the supremum for $\cS$, we define a $(d+1)$-dimensional vector $\bm{\Omega}$ with components
$\Omega_0 = 0$, $\forall k \in [d],$
\[
\Omega_k = \max\left( \sum_{i=1}^k\left[\bm{p}^{(1)}\right]^\downarrow_i,\; \cdots,\;
                \sum_{i=1}^k\left[\bm{p}^{(N)}\right]^\downarrow_i \right).
\]
The desired supremum $\bm{\omega}^{\sup}$ is then given by
\begin{equation}\label{eq:sup-omega}
\forall k \in [d],\; \omega^{\sup}_k = \Omega_k  - \Omega_{k-1}.
\end{equation}
The construction given in Eq.~\eqref{eq:sup-omega} guarantees that 
$\bm{\omega}^{\sup}$ majorizes every element of the set $\cS$, but $\bm{\omega}^{\sup}$ does not necessarily
appear in a descending order and may, therefore, fails to be majorized by other vectors with the same property.
In such case, we must perform a ``flattening'' process. This process starts with $\bm{\omega}^{\sup}$
obtained in Eq.~\eqref{eq:sup-omega}, and for every pair of components violating the descending order,
say, $\omega^{\sup}_k < \omega^{\sup}_{k+1}$, replaces the pair by their mean such that 
the updated two elements 
are $\widehat{\omega}^{\sup}_k = \widehat{\omega}^{\sup}_{k+1} = (\omega^{\sup}_k + \omega^{\sup}_{k+1})/2$.
This process continues until a descending vector corresponding to the supremum is obtained.

\subsection{Direct-sum majorization uncertainty}\label{sec:direct-sum-bound}

We now briefly introduce the uncertainty relation characterized by direct-sum majorization relation.
We remark that the results summarized here is originally presented in~\cite{rudnicki2014strong}.

Let $\cH$ be a $d$-dimensional Hilbert space. Denote by $\density{\cH}$ the set of quantum states in $\cH$.
Let $\bX$ and $\bZ$ be two rank-one projective observables, and $\rho$ be a state on $\cH$.
Assume the spectral decompositions of $\bX$ and $\bZ$ are given by
\[
\bX = \sum_{i=1}^d \alpha_i\proj{x_i},\quad
\bZ = \sum_{j=1}^d \beta_j\proj{z_j},
\]
where $\{\ket{x_i}\}$ and $\{\ket{z_j}\}$ are the eigenstates of $\bX$ and $\bZ$, respectively.
By measuring $\rho$, $\bX$ induces a probability distribution given by
\begin{equation}\label{eq:p_x}
\bm{p}\left(\bX\vert\rho\right) = \left(p_1, \cdots, p_{d}\right), \quad  
p_i = \bra{x_i}\rho\ket{x_i}.
\end{equation}
Similarly, $\bZ$ induces a probability distribution given by
\[
\bm{q}\left(\bZ\vert\rho\right) = \left(q_1, \cdots, q_{d}\right), \quad  
q_j = \bra{z_j}\rho\ket{z_j}.
\]

We are interested in the uncertainty relation induced by these two observables. 
In~\cite{rudnicki2014strong} the direct-sum majorization approach
is used to is to characterize the uncertainty about $\bm{p}(\bX\vert\rho)$ and $\bm{q}(\bZ\vert\rho)$:
\begin{equation}\label{eq:majorization-uncertainty}
\forall \rho \in \density{\cH},\quad
\bm{p}(\bX\vert\rho) \oplus \bm{q}(\bZ\vert\rho)
\prec \bm{\omega}^{\bX\oplus\bZ},
\end{equation}
where $\bm{\omega}^{\bX\oplus\bZ}$ is a $2d$-dimensional vector independent of $\rho$ which can be
explicitly calculated from observables $\bX$ and $\bZ$.
Intuitively, $\bm{\omega}^{\bX\oplus\bZ}$ is the supremum vector of the set
\[
\cS = \Big\{ \bm{p}(\bX\vert\rho) \oplus \bm{q}(\bZ\vert\rho): \rho \in \density{\cH}  \Big\}.
\]

Now we show how to compute $\bm{\omega}^{\bX\oplus\bZ}$ analytically.
From the definitions of $\bm{p}$ and $\bm{q}$, we can see that only the eigenstates of $\bX$ and $\bZ$
matter. We define a $d \times d$ unitary operator $U$ whose elements are given by
$U_{ij} = \langle x_i \vert z_j \rangle$.
$U$ is known as the \textit{overlapping matrix} as it characterizes the overlap of the two orthonormal bases. 
For each $k\in[d]$, let $\mathcal{SUB}(U,k)$ be the set of submatrices of class $k$ of $U$ defined as
\begin{multline}
\mathcal{SUB}(U,k) = 
\Big\{ M : \text{$M$ is a submatrix of $U$ satisfying~} \\ \sharp\opn{col}(M) + \sharp\opn{row}(M) = k+1 \Big\}.
\end{multline}
The symbols $\sharp\opn{col}(M)$ and $\sharp\opn{row}(M)$ denote the number of columns and rows of matrix
$M$, respectively. Based on the concept of submatrices,
we define the following set of coefficients, which is important in computing $\bm{\omega}^{\bX\oplus\bZ}$:
\begin{equation}\label{eq:s-k}
s_k = \max\Big\{\operatorNorm{M}: M \in \mathcal{SUB}(U,k)\Big\},
\end{equation}
where $\operatorNorm{M}$ is the operator norm of $M$,
and the maximum is optimized over all submatrices of class $k$.
By construction we have $c_1 = s_1 \leq \cdots \leq s_d = 1$.
In~\cite{rudnicki2014strong} it is proved that
$\bm{\omega}^{\bX\oplus\bZ} = \{1\}\oplus \bm{s}$, where $\bm{s}$ is given by
\[
\bm{s} = \left(s_1, s_2-s_1, \cdots, s_d - s_{d-1}, 0, \cdots, 0\right).
\]
We append $d-1$ $0$s to make $\bm{\omega}^{\bX\oplus\bZ}$ a $2d$-dimensional vector.
We remark that the vector $\bm{s}$ is not necessarily sorted in descending order,
but we can use the ``flattening'' process described in Sec.~\ref{eq:majorization} to make it descending ordered.
In words, the direct-sum majorization uncertainty relation can be summarized in the following theorem.
\begin{theorem}[\cite{rudnicki2014strong}]\label{thm:rudnicki2014strong}
Let $\bX$ and $\bZ$ be two rank-one projective observables on $\cH$ 
whose corresponding overlapping matrix is $U$. For any state $\rho\in \density{\cH}$, it holds that
$\bm{p}(\bX\vert\rho)\oplus\bm{q}(\bZ\vert\rho) \prec \bm{\omega}^{\bX\oplus\bZ} = \{1\} \oplus \bm{s}$.
\end{theorem}

\section{Entanglement detection}\label{sec:detection}

An entanglement detector decides whether a given bipartite state is separable by providing
a condition that is satisfied by all separable states, and if violated, witnesses entanglement.
In this section, we design a detection method 
based on the direct-sum majorization bound described in Sec.~\ref{sec:direct-sum-bound}.
As majorization relations, our detector actually provides a set of conditions whose number will
grow with the dimension of the state. We first describe a majorization bound
for all separable states. Then we show how this bound serves as a detector.
In the end, we illustrate by some examples how well the detector works.

\subsection{Majorization bounds}

If an observable $\bX$ is degenerate, the definition of $\bm{p}(\bX\vert\rho)$, given in Eq.~\eqref{eq:p_x},
is not unique, since the spectral decomposition is not unique. By combining eigenstates with the same eigenvalue,
however, there exists a unique spectral decomposition of the form $\bX = \sum_i \lambda_i P_i$,
with $\lambda_i\neq\lambda_{i^\prime}$ for $i \neq i^\prime$ and $P_i$ are orthogonal projectors of maximal
rank~\cite{guhne2004entropic}. Under this convention, we define for degenerate observable $\bX$ the distribution
$p_i = \tr\left[ P_i\rho\right]$. Our entanglement detection method relies on the degeneracy properties of the product
observables on bipartite systems. It is possible that for two non-commuting observables $\bX_A$ and $\bX_B$, their
product $\bX_A\otimes\bX_B$ is degenerate. Consequently, it may happen that $\bX_A\otimes\bX_B$ and $\bZ_A\otimes\bZ_B$
have a common eigenstate, and this eigenstate is an entangled pure state. In such cases, the probabilities
$\bm{p}(\bX_A\otimes\bX_B\vert\rho)$ and $\bm{p}(\bZ_A\otimes\bZ_B\vert\rho)$ will reflect the stated difference and
may be capable of detecting entanglement. As an example, consider the Pauli Z operator $\sigma_z$ on system $A$ and
$B$. The product observable on $AB$ is given by $\sigma_z\otimes\sigma_z$. The spectral decomposition of
$\sigma_z\otimes\sigma_z$ is (under our convention)
\[
\sigma_z\otimes\sigma_z 
= \left(\proj{00} + \proj{11} \right) - \left(\proj{01} + \proj{10} \right),
\]
Similarly, we have $\sigma_x\otimes\sigma_x = P_+ - P_{-}$, where $P_+ = \proj{++} +
\proj{--}$ and $P_{-} = \proj{+-} + \proj{-+}$, $\ket{+}=(\ket{0}+\ket{1}/\sqrt{2}$,
$\ket{-}=(\ket{0}-\ket{1}/\sqrt{2}$. There exists no state $\rho_A$ that can result in certain outcomes for both
$\sigma_x$ and $\sigma_z$, because they do not commute. But there do exist an entangled state $\ket{\Psi}$ that can
give certain outcomes for both $\sigma_x\otimes\sigma_x$ and $\sigma_z\otimes\sigma_z$, as they commute. By the Schmidt
decomposition, they can be expressed in the same eigenbases which are possibly entangled.


Let $\bX_A$ and $\bX_B$ be two full rank observables on $A$ and $B$, respectively.
Assume their spectral decompositions are given by
\[
\bX_A = \sum_{i=1}^d \alpha_i \proj{x^A_i},\quad
\bX_B = \sum_{i=1}^d \beta_i \proj{x^B_i}.
\]
Performing the product observable $\bX_A\otimes\bX_B$ on a bipartite state $\rho_{AB}$, we obtain a joint distribution
\[
p(i,j) = \bra{x^A_ix^B_j}\rho\ket{x^A_ix^B_j}.
\]
As $\bX_A\otimes\bX_B$ might be degenerate, some elements $p(i,j)$ are grouped together since they belong to the same
eigenvalue. We denote by $\bm{p}(\bX_A\otimes\bX_B\vert\rho)$ the joint distribution after grouping. If we perform
local observables, we obtain marginal distributions $\bm{p}(\bX_A\vert\rho_A)$ and $\bm{p}(\bX_B\vert\rho_B)$. It is
proved in~\cite{guhne2004entropic} that the joint distribution of a product state is majorized by the distribution of
its marginal.
\begin{lemma}[\cite{guhne2004entropic}, Lemma 1]\label{lemma:guhne2004entropic}
Let $\rho=\rho_A\otimes\rho_B$ be a product state and let $\bX_A$ and $\bX_B$ be two observables on $A$ and $B$,
respectively. Then
\begin{align*}
\bm{p}\left(\bX_A\otimes\bX_B\vert\rho_A\otimes\rho_B\right) 
&\prec \bm{p}\left(\bX_A\vert\rho_A\right), \\
\bm{p}\left(\bX_A\otimes\bX_B\vert\rho_A\otimes\rho_B\right) 
&\prec  \bm{p}\left(\bX_B\vert\rho_B\right).
\end{align*}
\end{lemma}
\noindent
Intuitively, this is because for the product observable $\bX_A\otimes\bX_B$, its eigenstates are possibly
entangled, and thus product state gives uncertain outcomes, however it is possible that
the reduced state gives certain outcome for the corresponding local observable.

Now we consider the effect of several product observables.
Let $\bX_A$ and $\bZ_A$ be two observables on $A$, $\bX_B$ and $\bZ_B$ be two observables on $B$, respectively.
For arbitrary product state $\rho=\rho_A\otimes\rho_B$, we obtain from Lemma~\ref{lemma:guhne2004entropic} that
\begin{align*}
\bm{p}\left(\bX_A\otimes\bX_B\vert\rho_A\otimes\rho_B\right)
&\prec \bm{p}\left(\bX_A\vert\rho_A\right), \\
\bm{p}\left(\bZ_A\otimes\bZ_B\vert\rho_A\otimes\rho_B\right)
&\prec \bm{p}\left(\bZ_A\vert\rho_A\right).
\end{align*}
As the direct-sum operation preserves the majorization order~\cite{marshall2010inequalities}, we have
\begin{align}
&\quad\;  \bm{p}\left(\bX_A\otimes\bX_B\vert\rho_A\otimes\rho_B\right)
          \oplus\bm{p}\left(\bZ_A\otimes\bZ_B\vert\rho_A\otimes\rho_B\right)\notag \\
&\prec \bm{p}\left(\bX_A\vert\rho_A\right) \oplus \bm{p}\left(\bZ_A\vert\rho_A\right).\label{eq:4}
\end{align}
The RHS. of Eq.~\eqref{eq:4} is the direct-sum of two distributions.
By the virtue of Thm.~\ref{thm:rudnicki2014strong}, it holds that
\begin{equation}\label{eq:optimal-omega-product-ds}
\forall \rho_A\in\density{\cH_A},\; \bm{p}\left(\bX_A\vert\rho_A\right) \oplus \bm{p}\left(\bZ_A\vert\rho_A\right)
\prec \bm{\omega}^{\bX_A\oplus\bZ_A}.
\end{equation}
Combining Eq.~\eqref{eq:4} and Eq.~\eqref{eq:optimal-omega-product-ds}, 
we reach the following statement for arbitrary product states $\rho_A\otimes\rho_B$, one has
\begin{equation}\label{eq:product-ds}
\bm{p}\left(\bX_A\otimes\bX_B\vert\rho_A\otimes\rho_B\right)
\oplus \bm{p}\left(\bZ_A\otimes\bZ_B\vert\rho_A\otimes\rho_B\right)
\prec \bm{\omega}^{\bX_A\oplus\bZ_A}.
\end{equation}

The majorization relation derived in Eq.~\eqref{eq:product-ds} holds for product states.
Now we show that this relation actually holds for arbitrary separable states. 
We are actually interested in the optimal state that
majorizes all possible probability distributions $\bm{p}\left(\bX_A\otimes\bX_B\vert\rho\right) \oplus
\bm{p}\left(\bZ_A\otimes\bZ_B\vert\rho\right)$ induced by performing $\bX_A\otimes\bX_B$ and $\bZ_A\otimes\bZ_B$
on \textit{separable} states. Such a state can be defined as
\begin{multline}\label{eq:optimal-omega-separable-ds}
\bm{\omega}^{(\bX_A\bX_B)\oplus(\bZ_A\bZ_B)}_{\opn{SEP}}
\coloneqq \sup\Big\{ \bm{p}\left(\bX_A\otimes\bX_B\vert\rho\right) \\  \oplus
                    \bm{p}\left(\bZ_A\otimes\bZ_B\vert\rho\right):
                    \rho \in \SEP(\cH_A{:}\cH_B) \Big\},
\end{multline}
where $\SEP(\cH_A{:}\cH_B)$ is the set of separable states of bipartite space $\cH_A\otimes\cH_B$.
In Appx.~\ref{appx:pure-states} we prove that 
$\bm{\omega}^{(\bX_A\bX_B)\oplus(\bZ_A\bZ_B)}_{\opn{SEP}}$ can be achieved among pure
product states, and thus we reduce the optimization over all separable states
required in Eq.~\eqref{eq:optimal-omega-separable-ds} to the optimization over all pure product states:
\begin{multline*}
\bm{\omega}^{(\bX_A\bX_B)\oplus(\bZ_A\bZ_B)}_{\opn{SEP}}
= \sup\Big\{\bm{p}\left(\bX_A\otimes\bX_B\vert\phi\right)\\ \oplus \bm{p}\left(\bZ_A\otimes\bZ_B\vert\phi\right):
    \phi = \proj{\phi_A}\otimes\proj{\phi_B}\Big\}.
\end{multline*}
For an arbitrary separable state (be it pure or not) $\rho_{AB}$, it then holds that
\begin{align*}\label{eq:5}
\bm{p}\left(\bX_A\otimes\bX_B\vert\rho\right) \oplus \bm{p}\left(\bZ_A\otimes\bZ_B\vert\rho\right)
&\prec \bm{\omega}^{(\bX_A\bX_B)\oplus(\bZ_A\bZ_B)}_{\opn{SEP}} \\
&\prec \bm{\omega}^{\bX_A\oplus\bZ_A}.
\end{align*}
The first inequality follows from the definition of $\bm{\omega}_{\opn{SEP}}$,
while the second inequality follows from the fact that each element
of $\bm{\omega}_{\opn{SEP}}$ is achieved by some pure product state,
and which in turn be majorized by $\bm{\omega}^{\bX_A\oplus\bZ_A}$ as proved in Eq.~\eqref{eq:product-ds}.
To summarize, we have the following theorem.
\begin{theorem}\label{thm:majorization-ds}
Let $\bX_A\otimes\bX_B$ and $\bZ_A\otimes\bZ_B$ be two product observables.
For arbitrary separable state $\rho\in\density{\cH_A\otimes\cH_B}$, it holds that
\[
\bm{p}\left(\bX_A\otimes\bX_B\vert\rho\right) \oplus \bm{p}\left(\bZ_A\otimes\bZ_B\vert\rho\right) 
\prec \bm{\omega}^{\bX_A \oplus \bZ_A},
\]
where $\bm{\omega}^{\bX_A \oplus \bZ_A}$ is defined in Thm.~\ref{thm:rudnicki2014strong}.
Similarly, one has
\[
\bm{p}\left(\bX_A\otimes\bX_B\vert\rho\right) \oplus \bm{p}\left(\bZ_A\otimes\bZ_B\vert\rho\right) 
\prec \bm{\omega}^{\bX_B\oplus\bZ_B}.
\]
\end{theorem}

\subsection{The detection framework}

Thm.~\ref{thm:majorization-ds} states that $\bm{\omega}^{\bX_A\oplus\bZ_A}$ is a necessary
condition for separability and its violation signals the existence of entanglement.
This statement provides an operational method of entanglement detection.
Given a bipartite state $\rho\in\density{\cH_A\otimes\cH_B}$, we first calculate
the direct-sum probability distribution 
$\bm{p}\left(\bX_A\otimes\bX_B\vert\rho\right)\oplus\bm{p}\left(\bZ_A\otimes\bZ_B\vert\rho\right)$  
induced by the product observables $\bX_A\otimes\bX_B$ and $\bZ_A\otimes\bZ_B$. 
Then we investigate the majorization relation between it
and $\bm{\omega}^{\bX_A\oplus\bZ_A}$.
If $\bm{\omega}^{\bX_A\oplus\bZ_A}$ does not majorize the direct-sum distribution,
then we conclude that $\rho$ is entangled. However,
if $\bm{\omega}^{\bX_A\oplus\bZ_A}$ majorizes the distribution, we can say nothing
about $\rho$: it can be separable, it can also be entangled.

The proposed method is a collection of linear detectors. Indeed, Thm.~\ref{thm:majorization-ds} states the following
fact. For arbitrary $k \in [2d]$, one has 
{
\small
\[
\sum_{i=1}^k \Big\{
    \bm{p}\left(\bX_A\otimes\bX_B\vert\rho\right) \oplus \bm{p}\left(\bZ_A\otimes\bZ_B\vert\rho\right)
            \Big\}^\downarrow_i
\leq \sum_{i=1}^k \Big\{ \bm{\omega}^{\bX_B\oplus\bZ_B} \Big\}^\downarrow_i.
\]
}As the first and the last $d$ inequalities are trivial,
we have $d-1$ effective inequalities in total, thus $d-1$ linear detectors.
States that violate any of these inequalities will necessarily be entangled.

\section{Entanglement detection via many observables}\label{sec:many-observables}

The entanglement detection method described in Sec.~\ref{sec:direct-sum} makes use of two incompatible observables
on each part. In this section, we generalize this method to the case of many incompatible observables.

Tensor-product majorization based uncertainty relations for many observables was first studied
in~\cite{xiao2016strong}. Here we show their results can be extended to the direct-sum majorization based uncertainty
relations. Let $\rho$ be a quantum state and $\{\bX^{(l)}\}_{l\in[L]}$ be a set of $N$ observables on $\cH$, where
$[L]=\{1,\cdots, L\}$. Assume the spectral decomposition of $\bX^{(l)}$ is given by
\[
\bX^{(l)} = \sum_{i=1}^d \alpha^{(l)}_i\proj{x_i^{(l)}},
\]
where $\{\ket{x_i^{(l)}}\}$ are the eigenstates of $\bX^{(l)}$.
By measuring $\rho$, $\bX^{(l)}$ induces a probability distribution given by
\begin{equation*}
\bm{p}\left(\bX^{(l)}{\Big\vert}\rho\right) = \left(p_1, \cdots, p_{d}\right), \quad  
p_i = \bra{x_i^{(l)}}\rho\ket{x_i^{(l)}}.
\end{equation*}
The direct-sum majorization based uncertainty relations for this set of observables has the following form:
\begin{equation*}
\forall \rho \in \density{\cH},\quad
\bigoplus_{l=1}^L\bm{p}\left(\bX^{(l)}{\Big\vert}\rho\right) 
\prec \bm{\omega}^{\oplus_{l=1}^L\bX^{(l)}},
\end{equation*}
where $\bm{\omega}$ is a $Nd$-dimensional vector independent of $\rho$ which can be
explicitly calculated from observables $\bX^{(l)}$.
To compute $\bm{\omega}$, we define the following coefficients
\begin{equation}\label{eq:s-k-many}
  s_k = \max_{\sum_{l=1}^L S_l = k} \lambda_1\big[ U(S_1,\cdots, S_L) \big],
\end{equation}
where $\lambda_1(A)$ denotes the maximal singular value of $A$, 
and the terms $S_l$, $U(S_1,\cdots, S_L)$ are defined in~\cite{xiao2016strong}. 
The main differences between our definition of $s_k$ in 
Eq.~\eqref{eq:s-k-many} and the $s_k$ defined in Eq.~15 of~\cite{xiao2016strong} lie in that
\begin{enumerate}
\item In our definition~\refeq{eq:s-k-many}, $S_l \geq 0$; while in their definition, $S_l$ is strictly positive.
\item In our definition~\refeq{eq:s-k-many}, the optimization is over all $\{S_l\}$
      such that $\sum_{x=1}^L S_l = k$; while in their definition,
      the optimization is over all $\{S_x\}$ such that $\sum_{x=1}^L S_l = k+L-1$.
\end{enumerate}
These two differences guarantee that we can use $s_k$ to give upper bounds on the 
sum of the first $d$ terms of $\bm{\omega}$. With coefficients $\{s_k\}$,
we can derive a direct-sum majorization bound for many observables.
\begin{lemma}\label{lemma:many-observables}
Let $\{\bX^{(l)}\}$ be a set of $L$ observables on $\cH$. For any state $\rho$ in $\cH$, it holds that
\begin{align*}
\bigoplus_{l=1}^L\bm{p}\left(\bX^{(l)}{\Big\vert}\rho\right) 
&\prec \bm{\omega}^{\oplus_{l=1}^L\bX^{(l)}} \\
&= \left(s_1, s_2-s_1, \cdots, L - s_a, 0, \cdots, 0\right),   
\end{align*}
where $s_{a+1}$ is the first component such that $s_{a+1}=L$.
\end{lemma}
The proof of Lemma~\ref{lemma:many-observables} is almost the same as the proof of Theorem 1 in~\cite{xiao2016strong},
with the modified definition of $s_k$ substituted.
Lemma~\ref{lemma:many-observables} is a generalization of Thm.~\ref{thm:rudnicki2014strong}
to the case of many observables. In Sec.~\ref{sec:detection}, we showed how 
Thm.~\ref{thm:rudnicki2014strong} is used to construct an entanglement detector.
We can also use Lemma~\ref{lemma:many-observables} to design entanglement detectors,
with the help of many observables.
\begin{lemma}\label{lemma:majorization-ds-many}
Let $\{\bX_A^{(l)}\}$ and $\{\bX_B^{(l)}\}$ be two sets of $N$ observables on $\cH_A$ and $\cH_B$, respectively.
For arbitrary separable state $\rho_{AB}$, it holds that
\[
\bigoplus_{l=1}^L \bm{p}\left(\bX^{(l)}_A\otimes\bX^{(l)}_B{\Big\vert}\rho\right) 
\prec \bm{\omega}^{\oplus_{l=1}^L\bX_A^{(l)}}.
\]
Similarly, one has
\[
\bigoplus_{l=1}^L \bm{p}\left(\bX^{(l)}_A\otimes\bX^{(l)}_B{\Big\vert}\rho\right) 
\prec \bm{\omega}^{\oplus_{l=1}^L\bX_B^{(l)}}.
\]
\end{lemma}
The proof of Lemma~\ref{lemma:majorization-ds-many} is similar to that of Thm.~\ref{thm:majorization-ds}.
Lemma~\ref{lemma:majorization-ds-many} provides an operational method of entanglement detection, using many
observables. Given a bipartite state $\rho_{AB}$, we first calculate the probability distributions
$\bm{p}\left(\bX^{(l)}_A\otimes\bX^{(l)}_B{\Big\vert}\rho\right)$ induced by the product observables
$\bX^{(l)}_A\otimes\bX^{(l)}_B$. This can be done by sampling from the source multiple times and gather the
statistics. Then we investigate the majorization relation between it and $\bm{\omega}$. If
$\bm{\omega}$ does not majorize the direct-sum distribution, then we conclude that $\rho_{AB}$ is
entangled.
\section{Conclusions}\label{sec:conclusion}

In this paper, we have studied the relationship between direct-sum majorization formulation of uncertainty relations
and entanglement, for the case of two and many observables. We have designed entanglement detection methods based on
such a formulation. Our nonlinear detectors are inherently stronger than similar scalar conditions as they are
equivalent to and imply infinite classes of such scalar criteria. 
Our measurement-based entanglement detection methods are of practical importance, as they are
experimental friendly and relatively easy to implement. We hope the results presented here can stimulate further
investigations on the relations among uncertainty relations, majorization, and entanglement.

\textit{Acknowledgments.}
This work is supported by the National Natural Science Foundation of China (Grant No. 61300050) and the Chinese
National Natural Science Foundation of Innovation Team (Grant No. 61321491).

\appendix


\section{Bounds are found on pure product states}\label{appx:pure-states}

Our task here is to establish the fact that direct-sum majorization induced bound
$\bm{\omega}_{\opn{SEP}}$ (defined in Eq.~\eqref{eq:optimal-omega-separable-ds})
can be achieved among pure product states.
Let $\mu_l$ be the $l$-th component of $\bm{\omega}_{\opn{SEP}}$. Assume w.l.o.g. that $\mu_l$ is achieved by 
the separable state 
$\widehat{\rho} = \sum_{k}\lambda_k \proj{\phi_k^A} \otimes \proj{\phi_k^B}$,
where $\{\ket{\phi_k^A}\}$ and $\{\ket{\phi_k^B}\}$ are orthonormal bases of $A$ and $B$, respectively.
Denote by $\cI$ ($\cJ$) be subsets of distinct index pairs from $[d]\times[d]$, and by
$\vert\cI\vert$ ($\vert\cJ\vert$) the size (number of elements) of $\cI$ ($\cJ$).
We assume the two probability sequences achieving $\mu_l$ are given
by $\cI$ and $\cJ$ satisfying $\vert\cI\vert + \vert\cJ\vert = l$. That is,
\[
\mu_l = \sum_{(i,j)\in\cI} p(i,j)  + \sum_{(m,n)\in\cJ} q(m,n),
\]
where $\bm{p}$ and $\bm{q}$ are the joint distributions given by product observable $\bX_A\otimes\bX_B$
and $\bZ_A\otimes\bZ_B$, respectively. From the linearity of the trace function, we have
\begin{align*}
p(i,j) 
&= \bra{x^A_ix^B_j}\widehat{\rho}\ket{x^A_ix^B_j}
  = \sum_k \lambda_k \vert\langle x^A_i x^B_j \vert \phi_k^A \phi_k^B \rangle\vert^2, \\
q(m,n) 
&= \bra{z^A_mz^B_n}\widehat{\rho}\ket{z^A_mz^B_n}
  = \sum_k \lambda_k \vert\langle z^A_m z^B_n \vert \phi_k^A \phi_k^B \rangle\vert^2.
\end{align*}
Thus
\begin{widetext}  
\begin{align*}
\mu_l &= \sum_{(i,j)\in\cI} p(i,j)  + \sum_{j\in\cJ} q(m,n)
=  \sum_k \lambda_k
    \left(\sum_{(i,j)\in\cI}\vert\langle x^A_i x^B_j \vert \phi_k^A \phi_k^B \rangle\vert^2 
+ \sum_{(m,n)\in\cJ}\vert\langle z^A_m z^B_n \vert \phi_k^A \phi_k^B \rangle\vert^2\right) \\
&\leq \max_{\ket{\phi_k^A\phi_k^B}}
  \sum_{(i,j)\in\cI}\vert\langle x^A_i x^B_j \vert \phi_k^A \phi_k^B \rangle\vert^2
+ \sum_{(m,n)\in\cJ}\vert\langle z^A_m z^B_n \vert \phi_k^A \phi_k^B \rangle\vert^2.
\end{align*}
\end{widetext}
\noindent 
That is to say, if $\widehat{\rho}$ achieves $\mu_l$, then $\widehat{\rho}$ must be a pure product
state, otherwise we can find a pure state which gives larger $\mu_l$ by simply 
choosing the eigenstate of $\widehat{\rho}$ with the largest eigenvalue.


\end{document}